# Distribution of contact forces in a homogeneous granular material of identical spheres under triaxial compression

P. Evesque
Lab MSSMat, UMR 8579 CNRS, Ecole Centrale Paris
92295 CHATENAY-MALABRY, France, e-mail: evesque@mssmat.ecp.fr

**Abstract:**

*The study of the distribution $\rho(f)$ of contact forces F in a homogeneous isotropic disordered granular sample subject to uniform triaxial stress field is undertaken using a model where forces propagate and collide. Collisions occur at grain and obey given rules which allow satisfying local static equilibrium. Analogy with Boltzmann's equation of density evolution is drawn and used to derive the parameters that control the distribution $\rho_s(f)$ of contact forces F in the stationary state in case of a packing of mono-disperse spheres. Using symmetry argument and mean field approximation, it is found that stationarity is achieved when the density $\rho_s(f)$ of force can be written as the product of exponentials of quantities whose sums are preserved during collisions. This introduces 3 parameters in 2d and 6 in 3d which are the mean force components $\{F_{xo}, F_{yo}, F_{zo}\}$, and the mean torques of the force on a grain $\{M_{xo}, M_{yo}, M_{zo}\}$. Astonishingly, it seems that the theory cannot include distribution of contact orientation implicitly. Extension of the model is possible with some care to case of anisotropic packing.*

**Pacs # : 5.40 ; 45.70 ; 62.20 ; 83.70.Fn**

---

Consider a sample of granular matter made of rigid spheres of identical size; it is supposed to be homogeneous and disordered, and to be submitted to an uniform triaxial stress σ. The questions are: what is the distribution of the contact forces? How does the correlation between forces vary with distance? What is the structure of the force network?

This has been the topic of a series of experimental and/or theoretical papers recently [1-7], leading to intricate modelling; for instance, a parallel with the physics of glasses has been even drawn [7]. Since Dantu [8] and De Josselin de Jong [9] one knows that the force field looks quite disordered. Indeed the main problem encountered is the correlation between the forces; to solve it, a subjacent network is postulated and a law of redistribution of forces is assumed; the problem becomes so intricate that solutions are found only when using drastic simplification, which forgets to ensure local equilibrium for instance [1-3]. These solutions ends with an exponential distribution $\rho(f)=\exp\{-f/f_o\}$. A similar exponential trend is also observed experimentally.

On the other hand, a simple solution to the same problem has been proposed [10] in the case of an isotropic stress; it was based on a simple statistical argument, "à la Boltzmann", which supposes that the force distribution obeys a principle of maximum disorder with two constraints in order to take into account (i) that the applied stress σ is fixed and (ii) that the total number of contacts is known and constant. This leads to predict that the contact forces obey an exponential distribution, *i.e.* $\rho(f)=\exp\{-f/f_o\}$ where $\rho(f)$ is the probability of finding a contact force whose modulus is f. Few former works have proposed some similar approach [11-13]. Indeed, as mentioned in the previous paragraph, experimental data are compatible with this law; hence they are compatible with this simplified modelling, which does not care of local equilibrium. Does it mean that the exponential distribution satisfy spontaneously local equilibrium?

This is just what we want to demonstrate; this is true even in a more complicated case, i.e. under a triaxial uniform stress field, as far as the medium is homogeneous.

The present article takes a more general point of view, which starts from the notion of force propagation along a network, transforming a space coordinate into "time coordinate". It imposes rules which ensure local static equilibrium.

Then, the evolution process is transformed into a problem of collisions, which allows the force network to evolve; hence the force network evolves due to interactions between forces, and such interactions are viewed as collisions. To warrant the existence of the local mechanical equilibrium, the model shall impose specific "collisions" rules, with preserved quantities. An equation of evolution of the distribution of forces is then found, which looks like a Boltzmann's equation of evolution [14-16].

Then the case of a homogeneous material subject to a triaxial uniform stress is investigated. It is shown that its distribution shall be stationary within the present modelling, due to the hypotheses of homogeneity and uniformity.

Next, the exponential distribution of force is shown to be a stationary solution of the problem that satisfies conditions of local equilibrium. This leads the paper to propose a general form for the stress distribution under triaxial confinement, which depends on 3 parameters in 2d, *i.e.* $\{F_{xo},F_{yo}, M_{zo}\}$, and 6 parameters in 3d, *i.e.* $\{F_{xo},F_{yo}, F_{zo}, M_{xo}, M_{yo}, M_{zo}\}$, which are the mean force- and mean momentum- components.

An important corollary (or consequence) of this study is the following: if one finds an experimental distribution $\rho_{exp}(f)$ that is different from the $\rho_{stat}(f)$, then it probably means that this experimental distribution $\rho_{exp}(f)$ is not stationary and the stress field not uniform.

The paper is built following these four steps, which are developed in successive subsections.

## 1. Force-propagation and force-networking:

Let us consider a homogeneous granular material made of grains of typical size d and subject to a uniform stress field. As shown in Fig. 1, the problem of the distribution of contact forces in a granular material can be formulated in terms of stress [4, 8-13 and refs there in], or in terms of a force network that propagates with some random character [1,6]; both are possible.

Indeed any grain can be viewed as stressed through a series of contact forces; this is the first case (or model A); it corresponds to the classic view where all the forces acting on a grain are viewed as applied from the other grains; and the sum of all the forces and momentums applied to a considered grain is equal to zero, because of mechanical equilibrium. The modelling holds either at the grain level or at any larger scale, for which it becomes the properties of normal uniform stress field.

It is straightforward to transform this model A into a model of propagation of forces: The basic idea in this case is to decompose the system into parallel slices and to order them in a direction perpendicular to the cuts. Now, each cut (C) cuts a given set $\{g_i\}$ of grains. Each of these grains $\{g_i\}$ can be considered as a force transmitters (model B), with incoming forces (which are those ones from one side of the cut) and with outgoing ones (which are those ones from the other side).

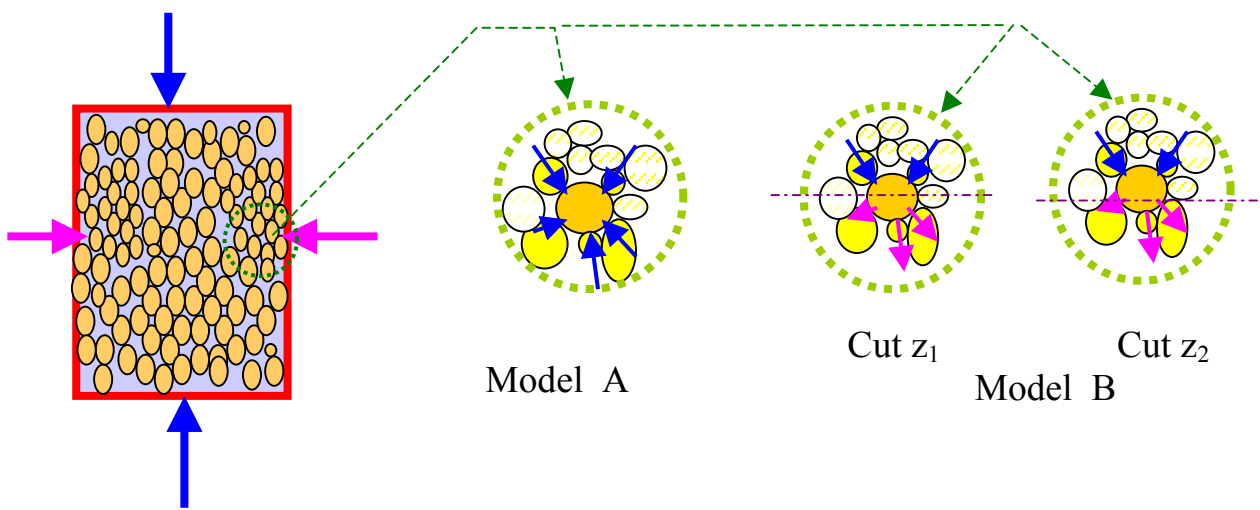

*Figure 1: Homogeneous packing of grains under homogeneous stress. Owing to the action of the surrounding grains, each grain is submitted to a series of forces. Model A allows to define the stress; Model B is issued from model A by inverting a series of forces (those forces which are below a given cut). This second model allows to make clear the mechanism of stress propagation. Equilibrium of each grain is ensured by $\Sigma_i F_i=0$ in model A on a grain, and by $\Sigma_i F_{i,\,incoming} = \Sigma_j F_{j,\,outgoing}$ in model B; this for each grain. The precise position z of the cut does not play an important role; it redistributes the exact distribution of incoming- and outgoing- forces.*

This is obtained as follow: the forces applied to a cut grain are separated into two categories, those which are from one side of the cut, *i.e.* (top), the others which are in the other side of the cut, *i.e.* (bottom). Then the second set of forces is inverted. Owing to this procedure, the first set can be viewed as the set of forces applied by the top part of the medium on the considered cut, and the second set as the set of forces applied by the cut on the bottom part of the medium. The first set will be called the set of incoming forces $\{f_{in,\,i}\}_C$, the second set the set of outgoing forces $\{f_{out,\,i}\}_C$.

• ***Rules for transmission inside the grains:*** In model B, equilibrium of a grain is obtained if the sum of its incoming forces is equal to the sum of its outgoing forces and if the sum of the momentums $\mathcal{M}_{in,i}$ generated by these incoming forces is equal to the sum of the momentums $\mathcal{M}_{out,i}$ generated by the outgoing forces, since both sets of forces generate momentums on the grain. This implies also to incorporate in the local description of incoming and outgoing forces the set of momentums, *i.e.* $\{f_{in,i}, \mathcal{M}_{in,i}\}_C$ and $\{f_{out,i}, \mathcal{M}_{out,i}\}_C$. The notation will be reduced to $\{f_{in,i}\}_C$ and $\{f_{out,i}\}_C$ for sake of simplicity, but it includes momentums.

• ***Sensitivity to the position of the cut :*** the "height" of the cut is somehow arbitrary; it influences only the number of forces in each category. It means that the network is quite sensitive to the precise position of each cut.

*Internal contacts:* One sees that some adjacent grains which are pertaining to the same cut have also common contacts; this is visible in the bottom diagram of Fig.2, for which forces are in green; this generates two opposite forces, both pertaining to the incoming set or to the outgoing set of forces. The momentums which are associated to them have opposite signs, but their modulus can be different if the two grains have different shapes and/or sizes. The existence of such internal contacts introduces some correlations between the set of incoming forces, and/or between the set of outgoing forces.

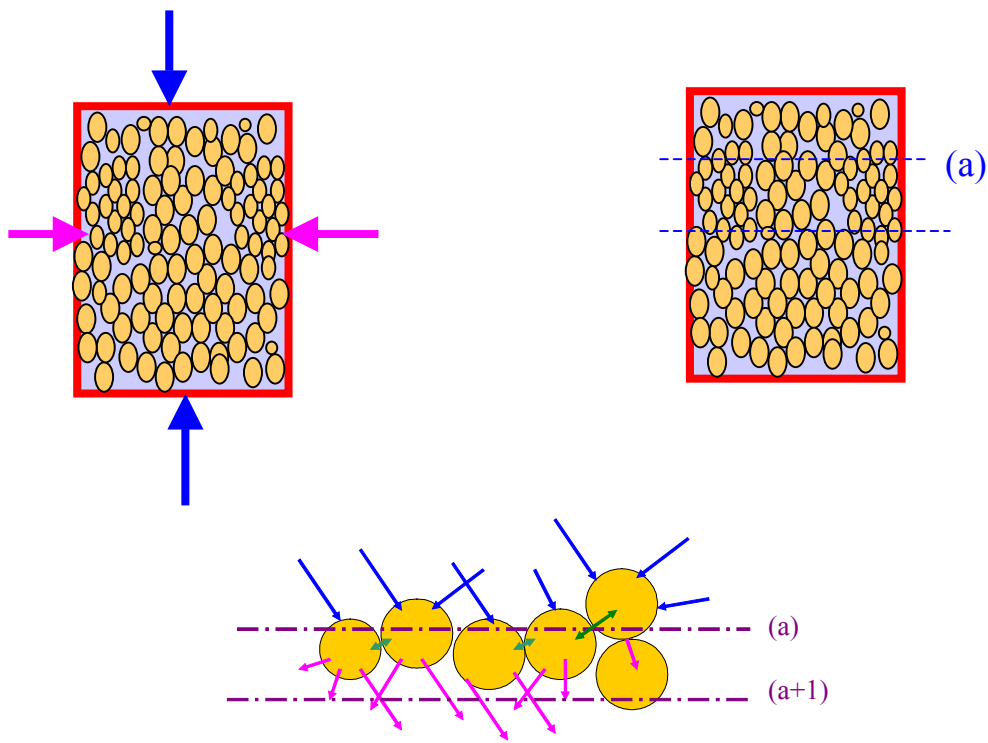

**Figure 2 : Left:** Packing of grains under triaxial stress. **Right:** two cuts at different height. **Bottom:** Enlargement of cut (a) to show the propagation of the force network between strata (a) and (a+1). The blue forces are the forces incoming at stage (a); the pink forces are force outgoing from strata (a); they are also forces incoming to strata (a+1). There exist also a series of green forces which are paired; both part of each green pair are either incoming or both outgoing forces, but each part of the pair pertain to a different grain. Furthermore, the sum of the two parts in each pair is zero, due to the law of action-to-reaction. Hence, green pairs introduce correlations between collision rules in adjacent grains.

• ***Rules for transmission to next grains:*** the set of forces $\{f_{out,i}\}_C$ can be viewed also as the set of forces applied to a new set of grains a little forward, since this set corresponds to next grains in the forward direction. This ensures the force to be transmitted. Hence, one can write $\{f_{out,i}\}_C = \{f_{in,i}\}_{C'}$, where C' is a new cut; this cut is no more flat; but the positions of the centres of the new grains it cut which are considered are located within a distance in between d/2 and 3d/2 from the initial cut C, since all these new grains shall touch a grain which is cut by C.

The problem of momentum transmission from one grain to the next one is some more tricky since the new momentum depends not only on the transmitted force, but also on the position, the shape and the size of the new grain. For instance, when considering spheres of identical sizes d, the momentums are just reversed, *i.e.* $\{\mathcal{M}_{out,i}\}_C = \{-\mathcal{M}_{in,i}\}_{C'}$, due to the position of the two centres, but when the spheres have different diameters d & d', the transmitted momentum is -d'/d time the initial one, *i.e.* $\mathcal{M}_{in,i,C'} = (-d'/d)\mathcal{M}_{out,i,C}$. Also the relation gets more intricate when grains are non spherical.

- ***Transmission from cut to cut:*** If one considers two successive cuts $C_n$ and $C_{n+1}$, they are separated by some distance ξ, filled with material; these two cuts are characterised by two different sets $[\{f_{in,i}\}_{C_n}, \{f_{out,i}\}_{C_n}]$ and $[\{f_{in,i}\}_{C_{n+1}}, \{f_{out,i}\}_{C_{n+1}}]$ and the medium transformed them into each other. So the medium in between can be viewed as a transformer which transforms either (i) the incoming set of forces and momentums in $C_n$ into the new incoming set of forces and momentums in $C_{n+1}$, *i.e.* $\{f_{in,i}\}_{C_n} \rightarrow \{f_{in,i}\}_{C_{n+1}}$, or (ii) the outgoing set of forces and momentums in $C_n$ into the new outgoing set of forces and momentums in $C_{n+1}$, *i.e.* $\{f_{out,i}\}_{C_n} \rightarrow \{f_{out,i}\}_{C_{n+1}}$, or (iii) the incoming set of forces and momentums in $C_n$ into the new outgoing set of forces and momentums in $C_{n+1}$, *i.e.* $\{f_{in,i}\}_{C_n} \rightarrow \{f_{out,i}\}_{C_{n+1}}$, or (iv) the outgoing set of forces and momentums in $C_n$ into the incoming set of forces and momentums in $C_{n+1}$, *i.e.* $\{f_{out,i}\}_{C_n} \rightarrow \{f_{in,i}\}_{C_{n+1}}$. One needs 3 of these 4 representations to define the complete details of the force transmission as soon as the distance ξ is non zero.

Of course, when ξ<d correlations exist between the different sets because some forces remain in common between them; this is true for $\{f_{in,i}\}_{C_n}$ & $\{f_{in,i}\}_{C_{n+1}}$, and between $\{f_{out,i}\}_{C_n}$ & $\{f_{out,i}\}_{C_{n+1}}$; some other forces of $\{f_{out,i}\}_{C_n}$ are in common with $\{f_{in,i}\}_{C_{n+1}}$; contrarily, there is ever no force in common between $\{f_{in,i}\}_{C_n}$ & $\{f_{out,i}\}_{C_{n+1}}$. Indeed, as soon as ξ≠0, (ξ <d), some old grains "disappear" and some new grains "appear" when passing from the cut $C_n$ to $C_{n+1}$; this generates a new set of forces. The collection of cut grains starts being completely different only when ξ>d.

In this case, direct correlation disappears between the sets of forces as soon as ξ>d. However correlations persist on larger scale due to conservation of macroscopic quantities. This is studied in the next paragraph.

- ***Large scale correlation :*** Indeed, correlation still persists at larger scale, due to conservation law. For instance, if one considers two cuts of size L² separated by ξ >d, with L>>ξ, one can decompose the medium in between the cuts in N cubes of size $ξ^3$, with N=(L/ξ)². Each facet of these cubes can be considered as a cut, so that one can proceed with facets perpendicular to the principal one as it has been proceeded with the normal one. So, including these new sets of incoming and outgoing forces and momentums, one can write macroscopic condition of equilibrium at scale ξ. This is the natural way to scale up the modelling.

Also, the equilibrium condition which has to be written in the case of a complex volume which contain few cubes and include internal facets shall not take care of the internal facets, because their contribution cancels. This is obvious in the case of a volume made of two adjacent cubes; the contribution of the internal facet has not to be taken into account since the facet pertains to the two cubes, and since it cuts grains in equilibrium; so both contributions counter-balance each other exactly. Following the same procedure, equilibrium condition can be written on larger parallelepiped volume containing many $(\lambda/\xi)^2$ cubes $\xi^3$; in the case $\lambda \gg \xi$ the effect of the outer lateral facets become negligible and equilibrium requires approximately the conservation of the sum of the forces on each face of size $\lambda^2$, *i.e.* it means normal rule of the transmission of mean stress.

A similar rule of conservation can be obtained for momentum transmission in the case of a packing of mono-disperse spheres, because momentum is correctly transmitted from sphere to sphere when spheres have the same size as stated above. However, it is no more true, when spheres of different sizes are mixed, or when grains have complex shapes; in this case transmission of momentums is more complex.

This point of view allows analysing the system as the propagation of forces in the direction perpendicular to the strata.

● ***Mapping the problem on a diffusion modelling:*** Let us define $x_1$ as the direction perpendicular to the cuts as in Fig. 2. Let us consider an inward contact force f ; it is located at a point of the network of contacts in the very vicinity of cut $x_1$. This force be viewed as propagating along $x_1$ with some evolution. A simple way to take into account this evolution is to consider that the force f diffuses also along $x_2$ and $x_3$; the diffusion is imposed by the interaction with the other forces which are applied to the same grain; the true interaction has been described already in the previous paragraphs, here the diffusion process is just an approximation; it ensures the conservation of the macroscopic quantity. Hence, the direction $x_1$ plays the role of the time and directions $x_2$ and $x_3$ the role of the space in this diffusion process; this point of view has been developed in [5].

If a collection of forces f located on a small area of the cut is now considered, with a density distribution ρ(f), it shall evolve also according to this diffusion; hence it shall obey:

$$\partial \rho(f)/\partial x_1 = D \left[ \partial^2 \rho(f)/\partial x_2^2 + \partial^2 \rho(f)/\partial x_3^2 \right] \tag{1}$$

where D is a diffusion coefficient, which is some kind of random variable that describes the mechanical rules of force transfer.

Indeed, the density distribution ρ(f) corresponding to this small surface of the given cut, say (a), located at $x_1$, is expected to look rather random and little correlated. In fact, correlations can be spontaneously generated (or destroyed) by the existence of "internal" contacts in the incoming (outgoing) set of forces; but these contacts are assumed to be few. Also pre-existing correlations can be transmitted partially; these correlations were pre-existing at earlier stage, say at $x_1-\delta x_1$ in (a-1), and are transmitted from direct or just-adjacent surfaces by the diffusion process. One expects this last contribution to be partially washed out because of the diffusion process.

However, it is obvious that the diffusion equation does not describe the exact mechanical rules which are obeyed by the force transfer. Hence, Eq. (1) is an approximation.

## 2. Collision formalism:

In this section we want to restate the problem with a different point of view. We keep in mind the view point of the evolution of the force network, which proceeds downward in Fig. 2, *i.e.* along the $x_1$ axis. But we reanalyse it from the viewpoint of collision: Indeed, any problem of diffusion or scattering can be viewed as a collision problem with a collision operator that characterise the local rules of transfer. In the present case, one has to describe that the evolution of the set of inward forces at stage (a). This set evolves due to the interaction between the inward forces themselves. The interactions occur within grains; hence the set of inward forces can be separated into disjoined subsets, each subset corresponding to a peculiar grain; hence each subset is made of the set of forces pointing inward the considered grain. Conversely, the sum of these subsets is the whole set of inward forces that correspond to cut (a).

Within this point of view, interactions between forces and momentums occur only within these subsets; they are localised on grain. These interactions produce the outward forces $\{f_{out,j,a}\}$ and the outward momentums $\{\mathcal{M}_{out,j,a}\}$. Then the outward forces transform directly into inward forces at cut (a+1), or on next grain, and so on, with the rule $f_{out,j,a}=f_{in,j,a+1}$. But the momentums are inversed, *i.e.* $\mathcal{M}_{in,j,a+1}=-\mathcal{M}_{out,j,a}$, when grains are spheres of equal size (the relationship between outward and inward momentums is simple only in the present case of equal spheres). This is the scheme of the evolution we want to study.

We note also from Fig. 2, that some inward (or outward) forces are paired (in green in Fig. 2); this occurs because they pertain to adjacent grains in contact and because these two grains are both cut by the cut (a). These pairs of forces shall be opposite because of the action-and-reaction law; hence they introduce some correlation between adjacent collision rules. But as they are not so many, we may expect their correlation effect is negligible. At least we will assume so hereafter.

So the propagation of the force network can be analysed in terms of a succession of collisions between inward forces that disappear and are replaced by the outward forces; in turn, these outward forces become the next incoming forces at the following grain, and so on. Within this scheme, collisions are localised on grains; they are described by collision rules. As the product of the collision is rather random set of force, these rules are rather random; however as the grains remain in static equilibrium the collisions obey some preserving rules which are momenta equal 0. In classic approach, this writes:

$$\Sigma_{\text{all forces } i \in \text{ same grain } g} \mathbf{f_{i,g}} = 0 \qquad (2.a)$$

$$\Sigma_{\text{all forces } i \in \text{ same grain } g} \mathbf{f_{i,g}} \wedge \mathbf{r_{i,g}} = 0 \qquad (2.b)$$

where bold characters are used for vectors, where $\wedge$ means vectorial product and where $\mathbf{r_{i,g}}$ is the vector connecting the centre of mass of grain g to the contact point where the force $f_{i,g}$ is applied.

However in the case of the present approach this shall be modified to account of the fact that incoming forces produce outgoing forces; hence Eq. (2) shall be rewritten as:

$$\Sigma_{\text{all incoming forces } i \in \text{ same grain } g} \mathbf{f_{i,g}} = \Sigma_{\text{all outgoing forces } j \in \text{ same grain } g} \mathbf{f_{j,g}} \qquad (3.a)$$

$$\Sigma_{\text{all incoming forces } i \in \text{ same grain } g} \mathbf{f_{i,g}} \wedge \mathbf{r_{i,g}} = \Sigma_{\text{all outgoing forces } j \in \text{ same grain } g} \mathbf{f_{j,g}} \wedge \mathbf{r_{i,g}} \qquad (3.b)$$

The evolution of the force network corresponds to a succession of propagation and collision. The collision matrix can be written:

$$\{\ldots, f_{i,\text{outgoing}}, \ldots\} = \underline{G_g} \{\ldots, f_{i,\text{incoming}}, \ldots\} \qquad (4a)$$

where $\underline{G_g}$ is a matrix whose coefficients depend strongly on the incoming forces and on the contact positions; but where $\underline{G_g}$ obeys Eq. (3) rules. Also $\underline{G_g}$ may contain some random value and some indeterminacy, when the system is hyperstatic.

Next step is the propagation to next grain; these rules can be written:

$$\{\ldots, f_{j,\text{incoming}}(x_1 + \alpha_{1ig}d, x_2 \pm \alpha_{2ig}d, x_3 \pm \alpha_{3ig}d), \ldots\} = \{\ldots, f_{i,\text{outgoing}}(x_1, x_2, x_3), \ldots\} \qquad (4.b)$$

$$\{\ldots, \mathcal{M}_{j,\text{incoming}}(x_1 + \alpha_{1ig}d, x_2 \pm \alpha_{2ig}d, x_3 \pm \alpha_{3ig}d), \ldots\} = \{\ldots, -\mathcal{M}_{i,\text{outgoing}}(x_1, x_2, x_3), \ldots\} \qquad (4.c)$$

where one notes the negative sign in Eq. (4c), which is due to the action of the force $f_{j,\text{incoming}}$ on the next grain as explained above. It demonstrate that the momentums are not preserved during this step; it means that if the force tends to make the grain $g_a$ rotating in one direction this force tends to make rotating the grain $g_{a+1}$ in the other direction, just because the centre of mass are symmetric compared to the point of application of the force. Also, d is the grain size in Eq. 4; here, it is assumed to be equal for all grains; and the $\alpha_{1ig}$ are parameters which depends on the geometrical configuration.

The sum of the rules Eq. 4a + Eqs 4(b&c) describe the complete collision-propagation rules, from a collision to another one.

So the objects which are colliding and propagating are the contact forces. It is worth noting at last that the number of forces is not preserved during collisions, because it depends on the number of contacts a grain has, which is a local random variable. This differs from numerous collision problems, where the number of particles that collide at a time is preserved; this is the case with liquid and gases. However, chemical reactions are well known examples which do not preserve particle numbers.

## 3. Solving the force propagation network using the formalism of the Boltzmann equation

Problems of collision and propagation are often solved using the formalism of the Boltzmann's equation. This is why it is tempting to do the same. A first step is to try and solve the stationary problem which implies $\partial\rho/\partial t=0$.

So let us consider a given distribution $\rho(F)$ of force F in the cut (a), one wants to estimate the evolution of this distribution at step (a+1). It is modified by a series of possible collisions. Some of them destroy the considered force; some other generates it from other forces; hence $\rho(F)$ at step (a) becomes $\rho(F)+\delta\rho(F)$ at step (a+1). And the stationary condition $\partial\rho/\partial t=0$ reads in the present case: $\delta\rho(F)/\delta x_1=0$, since $x_1$ plays the role of time.

Within the collision theory, the rate of disappearance of F is proportional to the probability of getting a collision with a contact force $F_{in}$ equal to F, the other contact forces spanning over the different possibilities. The rate of appearance of F is proportional to the probability of getting a collision such as one of the generated forces is equal to F.

So, if we consider that the force distribution is homogeneous and that no correlation exists between the forces in a same cut, we can write $\delta\rho(F)$ as the difference between the two terms.

$$\delta\rho(F)= \Sigma_{\text{configuration of g, such as one } F_{out}=F} \int G_g \prod_j[\rho(F_{out,j}) \, dF_{out,j}] - $$
$$- \Sigma_{\text{configuration of g}} \int G_g \, \rho(F_{in,k}=F) \prod_{j \neq k}[\rho(F_{in,j}) \, dF_{in,j}] \quad\quad (5)$$

And the stationary condition is obtained imposing $\delta\rho(F)= 0$.

Eq. (5) looks rather complicated. It works even when the number of contacts varies as a function of the grain and of the cut. But as we are looking for a homogeneous system in a stationary regime, we shall expect that $\delta\rho(F)=0$ in mean. As a matter of fact, as in each cut all the inward forces disappear to be replaced by outward forces of different values, Eq. (5) contains a large number of sums; so it is not obvious that the way Eq. (5) describes the problem is a simplified manner.

At last, but not least, the second (last) term of the right hand side of Eq. (5) describes the disappearance of F, while its first term describes the appearance of F. As a matter of fact the appearance of F occurs on the next grain; to evaluate its probability, one has to use the complete procedure described by Eqs. (4.a, b & c).

Hence it is a combination of two events, which are a collision and a propagation, which makes the equation rather complicated.

However, as it will be shown now, we may use the symmetry of the system and the preserving rules we have described already (Eq.3) to get simplification. In fact if we can enforce the use of preserving rules in the description of the system, we may introduce the correlations that are missing in most of the descriptions.

For instance, we may expect that the preserving rules (Eq. 3) are the main reason why local correlations exist, are generated and propagate; for instance, this ensures that large forces to propagate a while, because they can not sink into a single grain. On the other hand, we may expect* that the larger the sensitivity of the matrix $\underline{\underline{G}}_g$ to local conditions the better the decorrelation between adjacent forces.

## 4. Symmetry

As the sample is (i) homogeneous, (ii) is in static equilibrium and (iii) is subject to a homogeneous stress, we may consider the propagation of force in $+x_1$ direction or in the reverse direction, *i.e* $-x_1$. This shall not change the problem. Furthermore, looking at a slice does not allow to recognise the upward direction from the downward. So the two problems are equivalent. For instance if we consider a precise grain and cut, and a direction of propagation; this is called system 1; on this system we can reverse the propagating direction, this interchanges the sets of $\{.., F_{in} ,…\}$ with $\{.., F_{out} ,…\}$ and conversely; and this produces system 1'. But due to the symmetry of the sample and of force network, this new configuration can certainly correspond to a system 2, corresponding to a grain 2, with forces propagating in the direction of system 1; grain 2 is somewhere else in the cut. In this case, we can pair the two systems 1 and 2.

## 5. Stationary solution in case of homogeneous sample in a uniform stress field

Hence the stationary solution $\rho_s(F)$ is the one which satisfies that $\rho_s(F_{in})$ is produced by a set of $\rho_s(F_{out})$ and $\rho_s(F_{out})$ is produced by a set of $\rho_s(F_{in})$. In other words, the stationary problem is solved if one finds $\rho_s(F)$ that satisfies each pair:

$$G_g \{\prod_j[\rho_s(F_{out,j})] - \prod_k \rho_s(F_{in,k}) \} = 0 \qquad (6)$$

Solutions of Eq. (6) require that $\prod_j[\rho_s(F_{out,j})] = \prod_k[\rho_s(F_{in,k})]$ for a given collision. Hence it requires that $\prod_j[\rho_s(F_{out,j})]$ depends on the invariant of the collision only. These invariants are sums of variables I. Furthermore, the only solutions for $\rho_s = \rho_{stationary}$ are those which can be written as exponentials of linear combination of these variables.

$$\rho_{stationary}(F) = \exp\{-\Sigma_p a_p I_p\} \qquad (7)$$

Indeed, replacing [$\rho_s$ by Eq. (7) in Eq. (6) ensures to get 0. In Eq. (7) the $a_p$ are constant from collisions to collisions because it defines $\rho$, and the $I_p$ depends on the impact parameter $I_p$ which is then depending on and related to F. The impact parameters whose sums are preserved during collisions are the sum of forces and the sum of momentums. They are 3 in 2d, *i.e.* { $\Sigma_j F_{x,j}$, $\Sigma_j F_{y,j}$, $\Sigma_j (F_{x,j} r_{y,j} - F_{y,j} r_{x,j}) = \Sigma_j \mathcal{M}_{zj}$ } and 6 in 3d, *i.e.* { $\Sigma_j F_{x,j}$, $\Sigma_j F_{y,j}$, $\Sigma_j F_{z,j}$, $\Sigma_j (F_{y,j} r_{z,j} - F_{z,j} r_{y,j}) = \Sigma_j \mathcal{M}_{xj}$, $\Sigma_j (F_{z,j} r_{x,j} - F_{x,j} r_{z,j}) = \Sigma_j \mathcal{M}_{yj}$, $\Sigma_j (F_{x,j} r_{y,j} - F_{y,j} r_{x,j}) = \Sigma_j \mathcal{M}_{zj}$ }. Here {$F_x$, $F_y$, $F_z$} stand for the 3 components of forces and {$r_x$, $r_y$, $r_z$} stand for the 3 components of contact position in the mass centre of the considered grain.

The most general way to write $\rho_s$ that satisfies mean stationary condition at local stage, that is to say which satisfies Eq. (6) is then:

$$\rho_s(F) = \exp\{-a_x F_x - a_y F_y - b[F_x r_y - F_y r_x]\} \tag{8.a}$$

$$\rho_s(F) = \exp\{-a_x F_x - a_y F_y - a_z F_z - b_x [F_y r_z - F_z r_y] - b_y [F_z r_x - F_x r_z] - b_z [F_x r_y - F_y r_x]\} \tag{8.b}$$

We note that the last term in Eq. (8.a) and the 3 last ones in Eq. (8.b) depend on the normal at contact point when the grains are disks or spheres. So using the distribution of contact orientation may be adequate in such a case; however, this is no more exact as soon as the grain shape is more complex. Hence the use of contact orientation may still be approximating if the grains are round, but it is likely totally inadequate in case of non convex grains.

Parameters $a_x$, $a_y$, $a_z$ and $b_x$, $b_y$, $b_z$ are the inverses of the mean force components and mean momentums in direction x, y and z, *i.e.* $a_x=1/<F_x>$, $a_y=1/<F_y>$, $a_z=1/<F_z>$, $b_x=1/<\mathcal{M}_x>$, $b_y=1/<\mathcal{M}_y>$, $b_z=1/<\mathcal{M}_z>$; parameters $b_x$, $b_y$, $b_z$ are the inverse of the mean of momentums components exerted by a single contact force on a grain. Both series can be viewed as inversed of peculiar temperatures.

Parameters b/a depend on the distribution of contacts and on the orientation of the force at the contact on the grain. They depend on the friction angle $\varphi$ of the contacts since $\varphi$ limits the maximum torque-to-force ratio at limit of equilibrium; however, this may not be a simple relation; also the b/a values can be different from to the predicted limit in case of hyperstaticity.

## 6. Few Remarks

- The form of $\rho_s$ has been taken to impose grain equilibrium; then this form of $\rho_s$ **allows the preservation of forces and momentums** during collisions; in other word, $\rho_s$ **allows the preservation of forces and momentums in statics**.

- The counterpart of this choice has been to introduce a single density of contacts, that means that the density of contact does not depend on the orientation at this stage of the model. The model imposes a mean number of contacts; then it imposes a mean number of contacts per grain if all grains are identical

- The choice of $b_x$, $b_y$, $b_z$ controls the momentums; hence these coefficients controls the mean torque applied by a force; hence they depend on the mean friction forces.

- Anisotropy of contacts: In this model, the contacts seem to be oriented in all directions equivalently. However, the scales on $x_1$, $x_2$ and $x_3$ have not yet been defined at the moment; they were just supposed to be the same. But, the theory can be likely adapted to the case of an anisotropic model, with different length scales. For instance, let us defined different scales $\lambda_1$, $\lambda_2$, $\lambda_3$ for the three axes $x_1$, $x_2$, $x_3$ respectively, and the model applies directly to an anisotropic medium with an anisotropic distribution of contacts that depend on the contact orientation. As the length is different in each direction, this forces also to consider unit of forces as different in the 3 main directions, in order to ensure that the momentum balance is correctly written.

- The proposed model applies to the case of a single kind of grain only, and when the grains are spherical. When different sizes of grains are forming the sample, or when they are not spherical, the force transmission from grain to grain cannot be described by Eqs. (4b & 4c) as simply; but specific relations have to be written taking account of the probabilities of contacts between large-large, small-small, small-large & large-small pairs of grains and/or taking into account the distribution of possible orientations of the two grains. As noted earlier, one of the main difficulties comes from the non preservation of the momentums $\{\mathcal{M}_x, \mathcal{M}_y, \mathcal{M}_z\}$ through a contact in this case. However, the force remains transmitted correctly in accordance with the action-reaction principle. It results from this last preservation rule that some relation still exists between the different distributions in the stationary state. However things are made more complicated because of segregation, as sown here after.

- Link to segregation: Indeed, when different sizes of grains are forming the sample, one knows that segregation occurs. It implies that pairs of grains may not be uniformly distributed in the sample, so that large-small pairs of grains are oriented in a preferred direction, say $+x_1$, rather than in the opposite one, say $-x_1$. This makes the assumption of "time" reversal non valid anymore locally. This means a partial breaking of symmetry of the time reversal at least. This symmetry upon

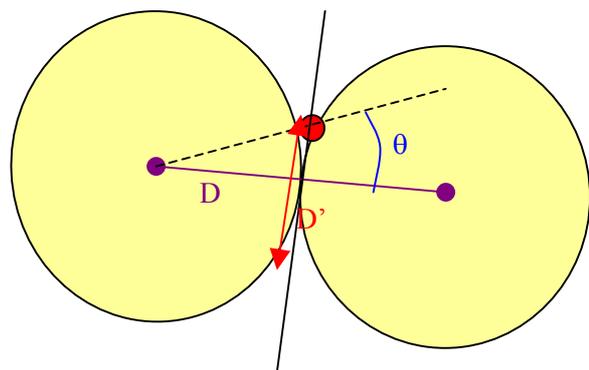

*Fig. 3 : When two large grains are in contact, their interaction excludes a large volume where no small grain can enters and make contact with one of these two grains.*

"time" can be even completely broken in case of gravity. This demonstrate that the analogues of Eqs. (4b &c) are much more complex. Also other problems arise, such as the effect of correlations between 3 grains as it is exemplified in Fig. 3: the distribution of contacts between 3 grains of different sizes present an excluded volume which screen part of the possible contact area of the larger grains….

## 7. Conclusion:

The aim of this paper is to investigate the distribution of contact forces in a granular material made of mono-disperse spheres and subject to a uniform triaxial stress. The basic idea of the present paper is to decompose the system in parallel slices and to order them in a direction perpendicular to the cuts; these cuts cut grains; then each cut grain is considered as a force transmitter, with incoming forces (which are those ones from one side of the cut) and the outgoing ones (which are those ones from the other side). This allows mapping the problem as a force propagation process. Then one can assume that the force obeys a given distribution $\rho(f)$ that evolves through the propagation process.

Then the paper uses a way rather similar to the one used by Boltzmann, when he developed the formalism of the Boltzmann's equation for gases, that describes the evolution of the probability density of the state of a single atom. In the case of perfect gases however, this approach can be used to demonstrate the principle of maximum disorder, by establishing the H theorem using conservation laws of energy and momentums. Then one can find the stationary solution of the distribution as the one which has the maximum disorder for a given mean energy. In the present case the stationarity cannot be demonstrated, but the difficulty is got round.

So the present paper pursues an analogy with the Boltzmann's approach, finding an equation of evolution of the distribution of contact forces in a granular medium subject to a given stress; for this, a space coordinate is transformed into "time". Then it analyses the evolution as a collision process, for which the conservation rules have been identified and ensure the stability of local equilibrium; then these rules are used to find a solution which is stationary. Unlike the Boltzmann theory of gas, one cannot demonstrate that the present force distribution shall be stationary; so it is one of the key hypotheses of this model. But it is a reasonable one, since assuming that both (i) the packing structure and (ii) the stress field are homogeneous is sufficient to imply that the contact force distribution is stationary. It means that the force distribution shall be the same on any surface perpendicular to a given direction, and shall depend only on this direction. Then conservation rules are used to enforce a set of possible distributions $\rho_{stat}(f)$ which remain invariant along propagation, *i.e.* stationary. This imposes an exponential distribution which depends on 6 parameters $\{a_x=1/<F_x>, a_y=1/<F_y>, a_z=1/<F_z>, b_x=1/<\mathcal{M}_x>, b_y=1/<\mathcal{M}_y>, b_z=1/<\mathcal{M}_z>\}$, which are the mean-force- and mean-momentum- components.

An important corollary (or consequence) of this study is the following: if one finds an experimental distribution $\rho_{exp}(f)$ that is different from the $\rho_{stat}(f)$, then it probably means that this experimental distribution $\rho_{exp}(f)$ is not stationary and the stress field not uniform.

*Acknowledgements:* CNES is thanked for partial funding.

The electronic arXiv.org version of this paper has been settled during a stay at the Kavli Institute of Theoretical Physics of the University of California at Santa Barbara (KITP-UCSB), in june 2005, supported in part by the National Science Fundation under Grant n° PHY99-07949.

*Poudres & Grains* can be found at :
http://www.mssmat.ecp.fr/rubrique.php3?id_rubrique=402